\documentclass[a4paper,12pt]{article}
\usepackage{amsfonts,slashed}
\usepackage{url}
\usepackage{latexsym}
\usepackage{amsfonts}
\usepackage{epsfig}
\usepackage{latexsym,amssymb}
  \usepackage{amsmath,amssymb,amsthm}
\usepackage{ifpdf}
\ifx\pdfoutput\undefined
   \pdffalse
   \usepackage{cite}
 \else
   \pdfoutput=1
   \pdftrue
  \usepackage[pdftex]{hyperref}
\fi

\numberwithin{equation}{section}
\def\be{\begin{equation}}
\def\ee{\end{equation}}
\def\ba{\begin{array}}
\def\ea{\end{array}}

\newcommand{\bea}{\begin{eqnarray}}
\newcommand{\eea}{\end{eqnarray}}

\newcommand{\bbox}{\lower.2ex\hbox{$\Box$}}

\def\bfone{\relax{\rm 1\kern-.35em 1}}
\def\bfzero{\relax{\rm I\kern-.18em 0}}

\begin{document}
\numberwithin{equation}{section}

\begin{center}
{\bf\LARGE On the dualization of Born-Infeld theories} \\
\vskip 2 cm
{\bf \large Laura Andrianopoli, Riccardo D'Auria and Mario Trigiante}
\vskip 8mm
 \end{center}
\noindent {\small{\it DISAT, Politecnico di Torino, Corso Duca
    degli Abruzzi 24, I-10129 Turin, Italy and Istituto Nazionale di
    Fisica Nucleare (INFN) Sezione di Torino, Italy}

\vskip 2 cm
\begin{center}
{\small {\bf Abstract}}\\

\end{center}
\vskip 1 cm
We construct a general Lagrangian, quadratic in the  field strengths of $n$ abelian gauge fields, which interpolates between  BI actions of n abelian vectors and actions, quadratic in the vector field-strengths, describing Maxwell fields coupled to non-dynamical scalars, in which the electric-magnetic duality symmetry is manifest. Depending on the choice of the parameters in the Lagrangian, the resulting BI actions may be inequivalent, exhibiting different duality groups. In particular we find, in our general setting, for different choices of the parameters, a ${\rm U}(n)$-invariant BI action, possibly related to the one in \cite{Aschieri:2008ns}, as well as the recently found $\mathcal{N}=2$ supersymmetric BI action \cite{Ferrara:2014oka}.

\vfill
\noindent {\small{\it
    E-mail:  \\
    {\tt
      laura.andrianopoli@polito.it}; \\
      {\tt riccardo.dauria@polito.it}; \\
      {\tt mario.trigiante@polito.it}}}
   \eject

\section{Introduction}

The Born-Infeld (BI) theory \cite{Born:1934gh} describes a non-linear electrodynamics in four dimensional space-time  enjoying remarkable features, among which electric-magnetic duality symmetry.
Such a peculiarity, which has been generalized to the case of $n$ abelian field strengths, where the duality group is contained in ${\rm Sp}(2n,\, \mathbb{R})$ \cite{Aschieri:1999jr},\cite{Aschieri:2000dx},\cite{Aschieri:2008ns}, hints to a connection of BI with extended supersymmetric theories, which also have the electric-magnetic duality invariance  \cite{Gaillard:1981rj} as a characteristic property. The supersymmetric version of the BI Lagrangian was constructed in \cite{Deser:1980ck},\cite{Cecotti:1986gb}, while in \cite{Casalbuoni:1988xh},\cite{Bagger:1996wp},\cite{Komargodski:2009rz} it  was identified as the invariant action of the Goldstone multiplet in a $\mathcal{N}=2$ supersymmetric theory spontaneously broken to $\mathcal{N}=1$. Recently, the results of \cite{Bagger:1996wp} have been generalized to the case of $n$ vector multiplets in $\mathcal{N}=2$ supersymmetry \cite{Ferrara:2014oka,Ferrara:2014nwa}, with explicit solutions for the case $n=2$ and $n=3$.

In this letter we provide a linear (in the squared field strengths) realization of the bosonic BI Lagrangian in terms  of a redundant Lagrangian containing two  couples of non dynamical scalars. The classical BI Lagrangian is recovered solving the field-equation constraints when varying our Lagrangian with respect to one of the two couples of scalars, while variation with respect to the other couple of Lagrange multipliers leads to a version of linear electromagnetism with generalized (scalar dependent) couplings and a positive scalar potential, in which the duality symmetry is manifest.
Remarkably, the properties of the resulting theory fit very well with
the bosonic sector of the $\mathcal{N}=2$ supersymmetric Lagrangian for a vector multiplet in the presence of a complex Fayet-Iliopoulos term, in the limit where the masses of the scalar sector are dominant with respect to their kinetic term. By appropriate choice of the normalization of the fields, we recover indeed, in a component form, the results of \cite{Bagger:1996wp}.

Let us remark that in our approach the possibility of dualization to BI is due to the presence of a scalar function $f(\Lambda)\propto \sqrt{1+\Lambda}$,  $\Lambda$ being one of the Lagrange multipliers. After implementing the proper normalization of the fields corresponding to the supersymmetric case, the coefficient in front of $f(\Lambda)$ turns out to be twice the product of an electric and a magnetic charge. In the absence of either the electric or the magnetic charge, our Lagrangian would reduce to linear electrodynamics coupled to scalars and it would not be able to implement the dualization to BI.
On the other hand, the need for both electric and magnetic charges is in fact a necessary condition for partial supersymmetry breaking $\mathcal{N}=2\to \mathcal{N}=1$, as shown in \cite{Antoniadis:1995vb,Rocek:1997hi}. Our formalism, recalling the results in \cite{Bagger:1996wp},  makes the relation between partial supersymmetry breaking and BI manifest.
Not surprisingly, the presence of $f(\Lambda)$ in our Lagrangian is also necessary to obtain, in the other version of the theory, a scalar potential  manifestly invariant under electric-magnetic duality symmetry.

In our framework, the generalization to more than one vector fields, at the purely bosonic level, is straightforward by promoting scalar fields to matrices. We write a general Lagrangian which also include some constant  matrices $\eta^{IJ}$, $\tilde \eta_{IJ}$.
In the generic case where $\eta^{IJ}$, $\tilde \eta_{IJ}$ are invertible, the extension of our approach to any number of vectors is straightforward  and leads to the definition of an abelian multi-field BI action which comprises, for a suitable choice of parameters, a $U(n)$-invariant BI action, possibly related to the one of \cite{Aschieri:2008ns} in the absence of extra scalar fields.
However, we show that we can relax the invertibility condition on the two constant matrices $\eta^{IJ}$, $\tilde \eta_{IJ}$, allowing for an $\mathcal{N}\geq 2$ supersymmetric extension. For specific choices of $\eta^{IJ}$, $\tilde \eta_{IJ}$ in terms of the electric and magnetic Fayet-Iliopoulos charges we reproduce the $\mathcal{N}=2$ supersymmetric BI action found in \cite{Ferrara:2014oka}.
Therefore, we show that, starting from our unifying description, different choices
of the constant matrices $\eta^{IJ}$, $\tilde \eta_{IJ}$ may lead, upon integrating out the non-dynamical fields, to inequivalent theories which exhibit different global symmetries.

\section{Linear realization of the Born-Infeld Lagrangian}
Let us consider the Born-Infeld Lagrangian in four dimensions:
\begin{eqnarray}
{\mathcal L}& =&\frac 1\lambda \Bigl\{1- \sqrt{\left|\det\left[\eta_{\mu\nu}+\sqrt{\lambda } F_{\mu\nu}\right]\right|}\Bigr\}=\nonumber\\
&=& \frac 1\lambda \left(1- \sqrt{1+ \frac\lambda 2 F^2 -\frac{\lambda^2}{16} (F\tilde F)^2}\right)\,,\label{bi}
\end{eqnarray}
where $F_{\mu\nu}= \partial_\mu A_\nu - \partial_\nu A_\mu$ is an abelian  field strength, $\tilde F_{\mu\nu}= \frac 12 F^{\rho\sigma}\epsilon_{\mu\nu\rho\sigma} $ its Hodge dual and
\begin{eqnarray}
F^2 &\equiv& F_{\mu\nu}F^{\mu\nu}\,,\\
F\tilde F &\equiv& \frac 12 F_{\mu\nu}F_{\rho\sigma}\epsilon^{\mu\nu\rho\sigma}\,.
\end{eqnarray}
We are going to show that it can be written as the standard Lagrangian of a gauge field-strength in a theory whose field content is enlarged to include two couples of scalar fields which play the role of Lagrange multipliers ${\tilde g},{\tilde\theta},\Lambda,\Sigma$:
\begin{eqnarray}
{\mathcal L}' = \frac{\tilde g}{2\lambda}\left( \Lambda +\Sigma^2-\frac\lambda 2\,F^2\right) +{\tilde\theta} \left(\frac14 F\tilde F -\frac \Sigma \lambda\right) +\frac 1\lambda \left(1- \sqrt{1+ \Lambda }\right)\,.\label{linbi}
\end{eqnarray}
Indeed, variation of ${\mathcal L}'$ in (\ref{linbi}) with respect to   ${\tilde g},{\tilde\theta}$:
\begin{eqnarray}
\frac{\delta {\mathcal L}'}{\delta {\tilde g}}=0 &\Rightarrow&  \Lambda   = \frac \lambda 2 F^2 -\Sigma^2\,,\\
\frac{\delta {\mathcal L}'}{\delta {\tilde\theta}}=0 &\Rightarrow& \Sigma = \frac\lambda 4 F\tilde F\,,
\end{eqnarray}
 yields the BI  Lagrangian (\ref{bi}), while variation with respect to  $\Lambda ,\Sigma$ allows to express them in terms of ${\tilde g},{\tilde\theta}$:
\begin{eqnarray}
\frac{\delta {\mathcal L}'}{\delta \Lambda}=0 &\Rightarrow& \bar \Lambda = {\tilde g}^{-2}-1\,,\\
\frac{\delta {\mathcal L}'}{\delta \Sigma}=0 &\Rightarrow& \bar \Sigma = \frac{\tilde\theta} {\tilde g} \,,
\end{eqnarray}
leading to the ``dual'' expression
\begin{eqnarray}
{\mathcal L}' = -\frac{{\tilde g}}{4}F^2 +\frac {\tilde\theta}4  F\tilde F -\mathcal V({\tilde g},{\tilde\theta})\,,\label{linbisusy}
\end{eqnarray}
where
\begin{eqnarray}
{\mathcal V}({\tilde g},{\tilde\theta}) &=& -\frac 1\lambda\left[ \frac {\tilde g}2 (\Lambda +\Sigma^2)-{\tilde\theta}\Sigma -\sqrt{1+\Lambda}+1\right]_{\Lambda=\bar\Lambda;\Sigma= \bar\Sigma}=\nonumber\\
&=& \frac 1{2\lambda}\left({\tilde g}+ {\tilde\theta}^2 {\tilde g}^{-1} + {\tilde g}^{-1} \right)-\frac 1\lambda \,.\label{pot}
\end{eqnarray}
Two properties of eq. (\ref{pot}) allow to embed eq. (\ref{linbisusy})  into a supersymmetric theory:
If we assume ${\tilde g}>0$, which gives the correct sign  to the gauge-field kinetic term in (\ref{linbisusy}), the potential $\mathcal V$ is positive definite (apart for an irrelevant additive constant). Furthermore, it can be written as
\begin{eqnarray}
{\mathcal V}({\tilde g},{\tilde\theta}) &=& \frac 1{2\lambda} {\rm Tr}[\mathcal M] -\frac 1\lambda \,,\label{potsusy}
\end{eqnarray}
where we introduced the matrix
\begin{eqnarray}
{\mathcal M}_{MN}[\tilde{g},\,\tilde{\theta}]=\begin{pmatrix}
{\tilde g}+ {\tilde\theta} {\tilde g}^{-1} {\tilde\theta} & -{\tilde\theta} {\tilde g}^{-1}\\
-{\tilde\theta} {\tilde g}^{-1} & {\tilde g}^{-1}
\end{pmatrix}\,,\label{Mgt}
\end{eqnarray}
which is familiar to supersymmetry and supergravity users, since it is the  symplectic matrix encoding the scalar-couplings to the gauge field-strengths in extended supersymmetric theories.

As shown below, (\ref{linbisusy}) can  be thought of as the bosonic sector of the Lagrangian of an $\mathcal{N}=2$ vector multiplet with a supersymmetry-breaking scalar potential, in a limit where the scalar-field kinetic term is negligible with respect to the potential term in the action. It will in fact turn out to coincide with the result of \cite{Ferrara:2014oka}. \par
The definition of  (\ref{potsusy}) in terms of an invariant quantity (the trace of the symplectic matrix ${\mathcal M}$) allows to define an extension of the  BI  Lagrangian to $n$ abelian vectors. This will be discussed in Section 3.

\subsection{Embedding of the 4D Born-Infeld action in $\mathcal{N}=2$ supersymmetry}
Let us consider an $\mathcal{N}=2$  vector multiplet, consisting of a gauge-vector $A_\mu$, a complex scalar $z$ and a couple of Majorana spinors $\lambda^A$ ($A=1,2)$.
The  bosonic  Lagrangian is
\begin{eqnarray}
{\mathcal L}= -\frac{g(z,\bar z)}4 F^2 + \frac{{\theta} (z,\bar z)}4 F \tilde F +  G_{z\bar z}\partial_\mu z\partial^\mu \bar z - {\mathcal V}_{\mathcal{N}=2}(z,\bar z)
\end{eqnarray}
where $g$ and $\theta$ are functions of the complex scalars $z,\,\bar{z}$ and  $G_{z\bar z}$ is the metric of the sigma-model. In this case, and in the absence of the hypermultiplet sector, the scalar potential ${\mathcal V}_{\mathcal{N}=2}$ is due to the presence of a (electric and magnetic)  FI  term ${\mathcal P}^x_M$ ($x=1,2,3$ is an $SU(2)$ index, $M=1,2$ is a symplectic one) such that  the supersymmetry transformation-law of the (chiral) gaugino acquires the shift $W^{z|AB}= {\mathrm{i}} (\sigma^x)^{AB}G^{z\bar z} \bar U_{\bar z}^M {\mathcal P}^x_M$, where $U^M_z= (f_z, h_z)$ is the symplectic section, $G^{z\bar z}$ the inverse of $G_{z\bar z}$, and
\begin{eqnarray}
{\mathcal V}_{\mathcal{N}=2}= \frac 12 W^{z|AB} G_{z\bar z} \bar W^{\bar z}_{AB}=\frac{1}{2} {\mathcal P}^x_M{\mathcal M}^{MN}
 {\mathcal P}^x_N\,,\label{VN2}
\end{eqnarray}
 where we used the special-geometry relation $U^M_z G^{z\bar z}\bar U^N_{\bar z}=\frac{1}{2}\left( {\mathcal M}^{MN}-{\mathrm i}\Omega^{MN}\right)$, having defined the symplectic metric $\Omega = \begin{pmatrix} 0&1\\-1&0\end{pmatrix}$.

The fermion shift generally fully breaks supersymmetry in the vacuum. However, by setting one of the three FI terms, say ${\mathcal P}^3$, to zero, thus breaking $SU(2)\to U(1)$, it is possible to preserve $\mathcal{N}=1$ supersymmetry. In this case,  considered in \cite{Ferrara:2014oka}, the spontaneously broken theory has a scalar potential  which  can be written in terms of a complex FI term ${\mathbf \mathcal P}=\frac{1}{\sqrt{2}}\,\Omega \,( {\mathcal P}^1 + {\mathrm i}{\mathcal P}^2)$ as:
\begin{eqnarray}
{\mathcal V}_{FPS} &=&  \bar{\mathbf\mathcal P}^M\left({\mathcal M}_{MN}+ {\mathrm i}\,\Omega_{MN}
\right) {\mathbf \mathcal P}^N = \nonumber\\
&=& m^2 \left[g +(\theta - \frac {e_1}m)^2g^{-1}\right]+ e_2^2 g^{-1} -2me_2 \,,\label{susy}
\end{eqnarray}
where, by fixing the $U(1)$ R-symmetry, we chose  ${\mathbf \mathcal P}^M= \begin{pmatrix}m\\ e_1 + {\mathrm i}\, e_2 \end{pmatrix}$. Let us denote by ${\mathcal L}_{FPS}$ the Lagrangian of \cite{Ferrara:2014oka}, with  scalar potential (\ref{susy}). The $\mathcal{N}=1$ scalar potential (\ref{susy}) differs from the $\mathcal{N}=2$ one by a constant additive term \cite{Antoniadis:1995vb,Ferrara:1995xi} depending on the product $m e_2$.\footnote{We thank Sergio Ferrara for enlightening clarifications on this point.} This extra term determines the vanishing of the $\mathcal{N}=1$ scalar potential on the supersymmetric vacuum.

In the vacuum, the scalar sector is completely fixed, while the gauge sector stays massless.

\vskip 5mm

Let us compare (\ref{susy}) with  (\ref{potsusy}).
For:
\begin{eqnarray}
\tilde g = \frac m{e_2}\,g\,,\quad
\tilde \theta = \frac m{e_2}\,\left(\theta -\frac{e_1}m\right)\,,\quad
\lambda=\frac{1}{2\,m^2}\,,\label{trans}
\end{eqnarray}
we find
\begin{equation}
{\mathcal V}_{FPS}(g,\theta)=\frac{e_2}{m}\,{\mathcal V}(\tilde g,\tilde \theta)\,,\qquad{\mathcal L}_{FPS}=\frac{e_2}{m}\,{\mathcal L}'+\frac{e_1}{4\,m} \,F\,\tilde{F}\,, \label{compare}
\end{equation}
showing that  (\ref{potsusy}) is  suitable to describe an $\mathcal{N}=2\to \mathcal{N}=1$ supersymmetric theory, if one reabsorbs the charges $m,e_1,e_2$  in the definition of the scalars $\tilde g,\tilde \theta$.

Restoring the auxiliary fields in $\mathcal{L}'$, as in (\ref{linbi}), we can rewrite  the Lagrangian in the following form
which dualizes the  BI  Lagrangian:
\begin{align}
\frac{e_2}m\,\mathcal{L}'&=-\frac{g}{4}\,F^2+\frac 1{4}\left({\theta}- \frac{e_1}{m}\right)\,F\tilde F+m^2\,g\,(\Lambda+\Sigma^2)-2\,m^2\left(\theta-\frac{e_1}{m}\right)\Sigma+2\,m e_2\left(1-\sqrt{1+\Lambda}\right) \,.
\label{linbisusy2}
\end{align}
The last contribution in (\ref{linbisusy2}),  needed for implementing the dualization into BI, requires $m\,e_2 >0$ (the same consistency condition was found in \cite{Ferrara:2014oka}). This shows that $e_2,m\neq  0$,  a necessary condition for partial supersymmetry breaking, is also  necessary for a supersymmetric Lagrangian to allow a non-linear realization of the gauge  sector.
In the supersymmetric vacuum, the scalars $\theta$, $g$ acquire a mass $M=\sqrt{2\frac{m^3}{e_2}}$. In the limit
$m\to \infty$ the scalar-field kinetic term is negligible and the scalars effectively behave as Lagrange multipliers.
  \par
 The transformation (\ref{trans}) amounts to a change of symplectic frame. Indeed, considering the ${\rm SL}(2,\mathbb{R})$-transformation
\begin{equation}
A^M{}_N=\frac{1}{\sqrt{e_2 m}}\,\left(\begin{matrix}m & 0\cr e_1 & e_2\end{matrix}\right)\,,\label{AA}
\end{equation}
under which the complex variable $z=\tilde{\theta}-i\,\tilde{g}$ transforms projectively $z \rightarrow z'= \frac 1 m (e_1+e_2 \,z)$, we have
\begin{equation}
\mathcal{M}[z',\bar{z}']=A^{-T}\mathcal{M}[z,\bar{z}]A^{-1}\label{traM}
\end{equation}
yielding  (\ref{compare}).

\section{Generalization to $n$ abelian vector fields}
Let us now discuss the generalization of the above construction to $n$ abelian gauge fields, and under which conditions it is possible to embed the bosonic lagrangian in a supersymmetric theory.

We found that the 1-vector  BI  Lagrangian (\ref{bi}) can be linearized  into (\ref{linbi}) with the help of two couples of auxiliary fields. The key, to obtain this result, was the introduction in (\ref{linbi}) of the function $f(\Lambda)\propto \sqrt{1+\Lambda}$, which reproduces  the BI Lagrangian for $\Lambda\rightarrow \Lambda(F,\tilde{F})=\frac\lambda 2 F^2 -\frac{\lambda^2}{16} (F\tilde F)^2$. We wish now to generalize  (\ref{linbi}) to $n$ vectors. The global symmetry of (\ref{linbi}) is manifest once we integrate-out   $\Sigma $ and $\Lambda$   and write the Lagrangian (modulo an overall factor and field redefinitions), in the form (\ref{linbisusy}) with scalar potential (\ref{potsusy}). From the latter, the ${\rm U}(1)$-duality invariance of the theory is manifest.

Generalizing (\ref{potsusy}) to $n$ vectors, it would be manifestly $\mathrm{U}(n)$-invariant, a distinctive feature of the dual BI theory which should then possibly be related to the action  of \cite{Aschieri:2008ns}.\par We shall actually generalize (\ref{potsusy}) to:
\begin{equation}
\mathcal{V}=\frac{1}{2\lambda}\,{\rm Tr}\left(N\mathcal{M}\right)+const.\,,\label{potN}
\end{equation}
where $N$ is a constant $2n\times 2n$ symmetric matrix. The global symmetries of the Maxwell equations close a group $G$ whose action on $\mathcal{M}$ amounts to symplectic transformations $A$:
$\mathcal{M}\rightarrow \mathcal{M}'=A^{-T}\,\mathcal{M}\,A^{-1}
$.
The  group $G$ is now contained in the intersection of the symplectic group with the invariance group of   $N$:
\begin{equation}
G\subset {\rm Sp}(2n,\,\mathbb{R})\cap{\rm Inv}(N)\,.
\end{equation}
If $N$ is positive-definite, its invariance group is ${\rm O}(2n)$ and $G\subset {\rm U}(n)$. We shall also discuss a limit where the matrix $N$ is singular, which is required if we wish to embed the model in a  supersymmetric context. Depending on the choice of $N$ and  on its invariance property, by integrating the auxiliary fields we shall end up with inequivalent BI Lagrangians. \par
Entering into the details of our construction,
we shall insist in demanding that the dualized BI theory have a scalar potential of the form (\ref{potN}). This fixes the function $f(\Lambda)$, thus providing a possible general definition for the n-vector duality-invariant BI Lagrangian.

Let us then introduce two couples of (matricial) auxiliary-fields $ g_{IJ}=  g_{JI}$, $ \theta_{IJ}=  \theta_{JI}$, and $\Lambda^{I J}$, $\Sigma^I_{\ J}$ ($I,J,\dots =1,\dots , n$), generalizing the fields $g,\,\theta, \Lambda,\,\Sigma$ of the $n=1$ case.
In particular, $ g_{IJ} >0$  and $ \theta_{IJ}$
 are the imaginary and real parts of a complex matrix $\mathcal{N}_{\Lambda\Sigma}\equiv \theta_{IJ}-i\,g_{IJ}$
 parametrizing the coset
$ \frac{{\rm Sp}(2n,\mathbb{R})}{{\rm U}(n)}$.
In terms of them we construct a symplectic, symmetric matrix $\mathcal{M}$ as in (\ref{Mgt})
\begin{eqnarray}
{\mathcal M}_{MN}[{g},\,{\theta}]=\begin{pmatrix}
{g}+ { \theta}\cdot {  g}^{-1} \cdot { \theta} & -{ \theta}\cdot {  g}^{-1}\\
-{  g}^{-1}\cdot  { \theta} & {  g}^{-1}
\end{pmatrix}\,,\label{Mgt2}
\end{eqnarray}
where now $M,N=1,\dots,  2n$. This matrix transforms, under the action of a symplectic transformation $A$ acting on $\mathcal{N}\rightarrow \mathcal{N}'$, as in (\ref{traM}):
\begin{equation}
{\mathcal M}[{g}',\,{\theta}']=A^{-T}\,{\mathcal M}[{g},\,{\theta}]\,A^{-1}\,.
\end{equation}

We  start from the  $n$-vector Lagrangian:
\begin{align}
{\mathcal L}' = \frac{ g_{IJ}}{2\lambda}\left( \Lambda ^{IJ} +(\Sigma \cdot \eta\cdot \Sigma^T)^{IJ}-\frac\lambda 2\,F^I_{\mu\nu}F^{J|\mu\nu}\right) +{\theta}'_{IJ}\left(\frac14 F^I\tilde F^J -\frac {(\Sigma\cdot \eta)^{IJ} }\lambda\right) +\frac 1\lambda \left(C- f(\Lambda )\right)\,,\label{nlinbi}
\end{align}
where
\begin{equation}
{\theta}'_{IJ}={\theta}_{IJ}-(\eta^{-1}\,\eta')_{IJ}\,
  \end{equation}
  and $\eta^{IJ}$, $\eta^{\prime I}{}_J$ are constant matrices, the former taken to be symmetric and, at this stage, non-singular. Moreover we also suppose $\eta^{-1}\,\eta'$ to be symmetric. The irrelevant constant $C$  is determined by convenience.\par
We shall determine $f(\Lambda)$ in order to obtain, upon integrating-out  $\Lambda,\,\Sigma$, a scalar potential of the form (\ref{potN}), for a certain symmetric matrix $N$   also assumed for the time being to be non-singular.  After eliminating $\Lambda^{IJ} $, $\Sigma^{I}_{\ J}$  through their field equations, the resulting Lagrangian should have the form
\begin{eqnarray}
{\mathcal L}' = -\frac{{ g}_{IJ}}{4}F^I_{\mu\nu}F^{J|\mu\nu} +\frac {\theta_{IJ}}4  F^I\tilde F^J -\mathcal V({ g},{\theta})\,,
\quad \mbox{with }\quad
{\mathcal V}({ g},{\theta}) =   \frac 1{2\lambda} \mathrm{Tr}\left({N\cdot\mathcal M}\right) -\frac C\lambda\label{npot}
\end{eqnarray}
where  we introduced  $N^{MN}\equiv \begin{pmatrix} \eta &\eta'\\ {\eta}^{\prime T}&\tilde \eta\end{pmatrix}$.
Through a symplectic transformation $S=\left(\begin{matrix}{\bf 1} & -\eta^{-1}\,\eta'\cr {\bf 0} & {\bf 1}\end{matrix}\right)$,  $N^{MN}$ can  be brought to the block-diagonal form
\begin{align}
N_D&= \left(\begin{matrix}\eta & {\bf 0}\cr {\bf 0} & \tilde{\eta}_0\end{matrix}\right)=S^T\,N\,S\,,\label{Stra}
\end{align}
 where $\tilde\eta_0\equiv \tilde\eta-{\eta}^{\prime T}\,\eta^{-1}\,{\eta}^{\prime}$,
provided $\eta^{-1}\eta'=\eta^{\prime T}\,\eta^{-1}$.
In the new frame  $\mathcal{M}$ reads:
\begin{align}
\mathcal{M}_0&= S^{-1}\,\mathcal{M}\,S^{-1\,T}=\left(\begin{matrix}g+\theta'\cdot g^{-1}\cdot \theta' & -\theta'\cdot g^{-1}
\cr -g^{-1}\cdot \theta' & g^{-1}\end{matrix}\right)\,.\label{MD}
\end{align}

Explicitly,
variation of (\ref{nlinbi}) with respect to $\Sigma \,,\Lambda$ gives
\begin{eqnarray}
 g\cdot  \Sigma \cdot\eta = \theta' \cdot \eta & \Rightarrow & \Sigma =  g^{-1}\cdot  \theta' + \omega\,\qquad \mbox{with $\omega$ : } \quad\omega\cdot \eta=0\,, \label{solvesigma}\\
 \frac{\partial f}{\partial \Lambda^{I J}}= \frac 12 g_{IJ} & \Rightarrow & f(\Lambda) =\frac 12\int  g_{IJ} d \Lambda ^{IJ}= \frac 12   g_{IJ} \Lambda ^{IJ} - \frac 12\int \Lambda^{IJ} d g_{IJ} \label{solveflambda}\,.
\end{eqnarray}
The $\omega$ term in (\ref{solvesigma}) is clearly trivial in the case we are considering now of a non-singular $\eta$. However, the same solution holds when   $\eta$ is singular (see next Section), in which case $\omega$ is non-vanishing.
 Substituting (\ref{solvesigma}) and (\ref{solveflambda}) into (\ref{nlinbi}), we get:
\begin{eqnarray}
{\mathcal V} &=& - \frac 1{\lambda}\left(\frac 12\int  \Lambda^{IJ} d g_{IJ} -\frac 12 {\rm Tr}( \eta \cdot\theta' \cdot  g^{-1}\cdot \theta') +1 \right)\,. \label{flambdapot}
\end{eqnarray}
By comparing (\ref{npot}) with (\ref{flambdapot}) we finally get
\begin{eqnarray}
\Lambda  &=& (g^{-1}\cdot \tilde\eta_0 \cdot g^{-1})  - \eta \,.\label{lambdasolved}
\end{eqnarray}
Using once more (\ref{solveflambda}), we find:
\begin{eqnarray}
f(\Lambda)=\mathrm{Tr}(g^{-1}\tilde{\eta}_0)= \mathrm{Tr}\left(\sqrt{\left(\eta + \Lambda\right)\cdot   \tilde\eta_0}\right)\,, \label{flambdasol}
\end{eqnarray}
where the matricial square root is intended as a solution in $g^{-1}\,\tilde{\eta}_0$ to the equation
\begin{equation}
\Lambda\,\tilde{\eta}_0  = (g^{-1}\cdot \tilde\eta_0)^2 - \eta\,\tilde{\eta}_0 \,,\label{lambdasolved2}
\end{equation}
subject to the condition $g_{IJ}>0$. This selects one out the possible solutions and defines a prescription for computing the square root.
For $n=1$ the Lagrangian (\ref{nlinbi}), using (\ref{flambdasol}), reduces to (\ref{linbisusy}), modulo an additive constant, if we set:
\begin{equation}
\eta=1\,\,,\,\,\,\eta'=\frac{e_1}{m}\,,\,\,\,\tilde{\eta}=\frac{e_1^2+e_2^2}{m^2}\,,\,\,\,\tilde{\eta}_0=\frac{e_2^2}{m^2}\,,\,\,\,
\lambda=\frac{1}{2m^2}\,\,,\,\,\,
f(\Lambda)= \frac{e_2}{m}\,\sqrt{1+\Lambda}\,.
\end{equation}
\par
With the above prescription for $f(\Lambda)$, by varying (\ref{nlinbi}) with respect to $ g$, $ \theta$, we obtain:
\begin{eqnarray}
{\mathcal L}& =& \frac 1\lambda \left\{C- \mathrm{Tr}\sqrt{(\eta\cdot \tilde \eta_0)^{I}_J+ \left[\frac\lambda 2 F_{\mu\nu}^I F^{K|\mu\nu} -\frac{\lambda^2}{16} \left(F\tilde F\cdot \eta^{-1}\cdot F\tilde F\right)^{IK}\right]\cdot (\tilde\eta_{0})_{KJ} }\ \right\}\label{nbi}
\end{eqnarray}
which gives a definition for the $n$-field generalization of the BI Lagrangian.
For convenience we  choose $C=\mathrm{ Tr}(\sqrt{\eta\cdot \tilde \eta_0})$.  In the case in which $N$ is the identity matrix, the model becomes ${\rm U}(n)$-invariant. However the relation between the above Lagrangian and the ${\rm U}(n)$-invariant BI model of \cite{Aschieri:2008ns}, besides the common duality invariance, is not apparent and deserves further investigation. We refrain here from addressing the issue of uniqueness of  the ${\rm U}(n)$-invariant BI model.

\subsection{A singular limit: The $\mathcal{N}=2$ supersymmetric case}

For $\mathcal{N}=2$ superymmetric theories with $n>1$ vector multiplets and complex   FI  terms $ {P}^M $
it is possible to write the scalar potential as \cite{Ferrara:2014oka}:
\begin{equation}
{\mathcal V}_{FPS}(z,\,\bar{z})=\bar{P}^M\,\mathcal{M}(g,\theta)_{MN} P^N +i\,\bar{P}^M\,\Omega_{MN} P^N\,,\label{sergpot}
\end{equation}
where $g_{IJ}=g_{IJ}(z,\,\bar{z}),\,\theta_{IJ}=\theta_{IJ}(z,\,\bar{z})$, depend on $n$ complex scalars $z^i$, and \footnote{In general  $P^M=(m_1^I+i\,m^I_2,\,e_{1I}+i\,e_{2I})$. However, using a ${\rm U}(n)$ transformation we can always set $m_2^I=0$.}
\begin{equation}
P^M=({\bf m},\,{\bf e}_{1}+i\,{\bf e}_{2})\,\,,\,\,\,{\bf m}\equiv (m^I)\,\,,\,\,\,{\bf e}_1\equiv (e_{1\,I})\,\,,\,\,\,{\bf e}_2\equiv (e_{2\,I})\,.
\end{equation}
Such a potential induces  partial supersymmetry  breaking, so that one of the two supersymmetries is realized non-linearly and the scalar fields become massive.
It can be cast in the  form (\ref{npot})
\begin{equation}
{\mathcal V}_{FPS}(g,\theta)=\frac{1}{2\lambda}\,{\rm Tr}\left({N}\,\mathcal{M}[g,\theta]\right)
-\frac{C}{\lambda}
\,,\label{genformV}
\end{equation}
by choosing
${N}^{MN}= 2\,\lambda\, P^{(M }\bar P^{N)}$,
that is:
\begin{align}
\eta=&2\,\lambda\,{\bf m}\,{\bf m}^T=2\,\lambda\,(m^I\,m^J)\,\,,\,\,\,\eta'=2\,\lambda\,{\bf m}\,{\bf e}_1^T=2\,\lambda\,(m^I\,e_{1\,J})\,\,,\,\,\,
\tilde{\eta}=2\lambda\,({\bf e}_1\,{\bf e}_1^T+{\bf e}_2{\bf e}_2^T)\,,\label{etasusy}\\
\lambda=&\frac{1}{2\,{\bf m}^T {\bf m}}\,\,,\,\,\,C=2\,\lambda\,{\bf m}^T\,{\bf e}_2\,.\nonumber
\end{align}
In this case $ N^{MN}$  has rank-2, and it is not invertible for $n>1$.

 This case can be included in the general analysis performed above as a singular limit. In particular all formulas up to Eq. (\ref{lambdasolved2}) apply also to this case. With respect to our previous analysis we have however the following important differences:
  \begin{itemize}\item{ Formula (\ref{nbi}) was derived by varying the Lagrangian with respect to $g_{IJ},\,\theta_{IJ}$ considered as independent fields. In a supersymmetric model, as emphasized above, the two matrices are not independent but are functions of the complex scalar fields $z^i$. In order to eliminate the auxiliary fields in favor of the field strengths, therefore, a different set of equations should be solved, see below;} \item{Being now $\eta$ singular, the diagonalization of the matrix $N$ is  effected by a different symplectic transformation of the form:
  \begin{equation}
  S=\left(\begin{matrix}{\bf 1} & {\bf s} \cr {\bf 0} & {\bf 1}\end{matrix}\right)\,\,,\,\,\,\,{\bf s}=-\frac{1}{({\bf m}^T {\bf e}_1)}\left(e_{1 I} e_{1 J}\right)\,.
  \end{equation}
  The diagonal matrix $N$ has the same form as in  (\ref{Stra}) with:
  \begin{equation}
  \tilde{\eta}_0=\tilde{\eta}+\eta^{\prime\,T}\,{\bf s}=2\,\lambda\,\left(e_{2 I} e_{2 J}\right)\,,
  \end{equation}
 and in the new frame $\mathcal{M}_0$ has the same form as in (\ref{MD}) with:
  \begin{equation}
  \theta'_{IJ}=\theta_{IJ}-\frac{1}{({\bf m}^T {\bf e}_1)}\,e_{1 I} e_{1 J}\,.
  \end{equation}
  }
 \end{itemize}

We then start from  (\ref{nlinbi}), with the definitions  (\ref{etasusy}), $f(\Lambda)$ given in (\ref{flambdasol}), and with the matrices $g,\,\theta$ intended as functions of $z^i$,  and implement the constraints (\ref{solvesigma}) and (\ref{lambdasolved}) in order to get, in the $m^I \to \infty$ limit,  the bosonic sector of the $\mathcal{N}=2\to \mathcal{N}=1$ Lagrangian $\mathcal{L}^{(0)}_{FPS}$:
\begin{equation}
\mathcal{L}^{(0)}_{FPS}(z,\bar{z},\,F)=-\frac{{\rm Tr}(F^T\,g\,F)}{4}+\frac{{\rm Tr}(F^T\,\theta\,\tilde{F})}{4}-{\mathcal V}_{FPS}(z,\,\bar{z})\,,\label{theabovelagrangian}
\end{equation}
where the kinetic term of the scalars $z^i$ was omitted since subleading for  $m^I \to \infty$. Computing (\ref{theabovelagrangian}) on the $\mathcal{N}=1$-solution to the field-equations for $z^i$,  one obtains the BI Lagrangian of \cite{Ferrara:2014oka}. Now, however, variation with respect to $\Lambda$ of  (\ref{nlinbi}), with $f(\Lambda)$ given by (\ref{flambdasol}), does not reproduce (\ref{solveflambda}). The reason is that  (\ref{lambdasolved}) can  no longer be inverted to express $g$ in terms of $\Lambda$, so that (\ref{flambdasol}) should be intended as describing $f(\Lambda)$ \emph{only} on the solution (\ref{lambdasolved}): $f(\Lambda(g))$.
However
 \begin{eqnarray}
\left.\frac{\partial\mathcal{L}'}{\partial \Lambda}\right\vert_0 \cdot\frac{\partial \Lambda }{\partial z^i}=0\,,\quad
\left.\frac{\partial\mathcal{L}'}{\partial \Sigma}\right\vert_0 \cdot\frac{\partial \Sigma }{\partial z^i}=0\label{flz} \end{eqnarray}
still hold, the zero-subscript meaning that the quantity is computed on the solutions $\Lambda(z,\bar{z}),\,\Sigma(z,\bar{z})$ given by (\ref{lambdasolved}), (\ref{solvesigma}) with $g,\,\theta$ intended as functions of $z,\,\bar{z}$.
Moreover we still have:
\begin{equation}
\mathcal{L}^{(0)}_{FPS}(z,\bar{z},\,F)=\mathcal{L}'(\Lambda(z,\bar{z}),\,\Sigma(z,\bar{z}),z,\bar{z},F)\,.\label{fn}
\end{equation}
Properties (\ref{flz}) and (\ref{fn}) are enough to guarantee that the field-equations for $z^i$ obtained from $\mathcal{L}^{(0)}_{FPS}(z,\bar{z},\,F)$ are equivalent to those obtained from $\mathcal{L}'$ once we write for $\Lambda $ and $\Sigma$ their values $\Lambda(z,\bar{z}),\,\Sigma(z,\bar{z})$:
\begin{align}
\frac{\partial \mathcal{L}^{(0)}_{FPS}}{\partial z^i}=\left.\frac{\partial \mathcal{L}'}{\Lambda^{IJ}}\right\vert_0\,\frac{\partial \Lambda^{IJ} }{\partial z^i}+\left.\frac{\partial \mathcal{L}'}{\Sigma^{IJ}}\right\vert_0\,\frac{\partial \Sigma^{IJ} }{\partial z^i}+\left.\frac{\partial \mathcal{L}'}{\partial z^i}\right\vert_0=\left.\frac{\partial \mathcal{L}'}{\partial z^i}\right\vert_0\,.
\end{align}
\emph{As a consequence, the $\mathcal{N}=2$ BI action of \cite{Ferrara:2014oka} can also be obtained  from $\mathcal{L}'$   solving the field-equations for $z^i$.}\par
The problem with the non-invertibility of  (\ref{lambdasolved}) can be circumvented by \emph{regularizing} $\mathcal{L}'$ as follows. We define
$\mathcal{L}'_\epsilon\equiv \left.\mathcal{L}'\right\vert_{\eta\rightarrow \eta^\epsilon\,,\,\,\tilde{\eta}_0\rightarrow \tilde{\eta}_0^\epsilon}\,,
$
where $ \eta^\epsilon$ and $\tilde{\eta}_0^\epsilon$ are now non-singular:
\begin{align}
\eta^\epsilon\equiv {\bf m}\,{\bf m}^T+\epsilon\,\sum_{\alpha=1}^{n-1}{\bf m}_\alpha\,{\bf m}_\alpha^T\,\,,\,\,\,{\bf m}^T\,{\bf m}_\alpha=0\,\,,\,\,\,{\bf m}_\alpha^T{\bf m}_\beta=\delta_{\alpha\beta}\,,\nonumber\\
\tilde{\eta}_0^\epsilon\equiv {\bf e}_2\,{\bf e}_2^T+\epsilon\,\sum_{\alpha=1}^{n-1}{\bf e}_{2\,\alpha}\,{\bf e}_{2\,\alpha}\,\,,\,\,\,{\bf e}_2^T\,{\bf e}_{2\,\alpha}=0\,\,,\,\,\,{\bf e}_{2\,\alpha}^T\,{\bf e}_{2\,\beta}=\delta_{\alpha\beta}\,.
\end{align}
The field-equations for $\Lambda,\,\Sigma$ obtained from $\mathcal{L}'_\epsilon$ are solved by  $\Lambda_\epsilon(z,\bar{z}),\,\Sigma_\epsilon(z,\bar{z})$ and $\mathcal{L}^{(0)}_{FPS}(z,\bar{z},\,F)$ is obtained in the singular limit:
\begin{equation}
\mathcal{L}^{(0)}_{FPS}(z,\bar{z},\,F)=\lim_{\epsilon\rightarrow 0}\mathcal{L}_\epsilon'(\Lambda_\epsilon(z,\bar{z}),\,\Sigma_\epsilon(z,\bar{z}),z,\bar{z},F)\,.
\end{equation}
This formal derivation does not affect the above conclusion about the resulting BI action.\par
The field-equations   for the scalars $z^i$ from $\mathcal{L}'$ are  conveniently written in the special-coordinate description of the scalar manifold ($z^i=X^I$):
\begin{align}
C_{IJK}\left[-\frac{i}{2\lambda}\,\left(\Lambda+(\Sigma\eta\Sigma^T)-\frac{\lambda}{2}\,FF\right)^{IJ}+
\left(\frac{F\tilde{F}}{4}-\frac{\Sigma\eta}{\lambda}\right)^{IJ}\right]=0\,,\label{eqsf}
\end{align}
where we defined $C_{IJK}=\partial_I\partial_J\partial_K F$, $F(X)$ being the holomorphic prepotential, $\partial_I\equiv \frac{\partial}{\partial X^I}$, and we  used the property $g_{IJ}={\rm Im}(\partial_I\partial_J F)$, $\theta_{IJ}={\rm Re}(\partial_I\partial_J F)$.\par
\paragraph{Adapting the auxiliary-field description to $\mathcal{N}=2$ notation.\\}
To make contact with the supersymmetry notation, we  write $\Lambda$ and $\Sigma$ in terms of $2n$ auxiliary fields $\hat{\Phi}_1^I$, $\Phi_2^I$:
\begin{equation}
\Lambda^{IJ}=2\,\lambda\,\hat{\Phi}_1^I\,\hat{\Phi}_1^J-\eta^{IJ}=2\,\lambda\,(\hat{\Phi}_1^I\,\hat{\Phi}_1^J-m^I\,m^J)\,,\,\,\,\Sigma^I{}_J m^J=\Phi_2^I\,.\label{auxi}
\end{equation}
We find:
\begin{align}
f(\Lambda(\hat{\Phi}_1))={\rm Tr}\sqrt{(\eta+\Lambda)\,\tilde{\eta}_0}={\rm Tr}\sqrt{2\lambda\,\hat{\Phi}_1\hat{\Phi}_1^T\,\tilde{\eta}_0}=2\lambda\,{\rm Tr}\sqrt{(\hat{\Phi}_1\,{\bf e}_2^T)^2}=2\lambda\,\hat{\Phi}_1^I\,e_{2\,I}\,.
\end{align}
The resulting Lagrangian $\mathcal{L}''$ in terms of $\hat{\Phi}_1,\Phi_2,z, F$ now reads:
\begin{align}
\mathcal{L}''(\hat{\Phi}_1,\,\Phi_2,\,z,\bar{z},F)=&g_{IJ}\left(\hat{\Phi}_1^I\hat{\Phi}_1^J-m^I\,m^J+\Phi_2^I\,\Phi_2^J-
\frac{1}{4}\,F^I\,F^J\right)+\nonumber\\
&+\left(\theta_{IJ}-\frac{e_{1 I} e_{1 J}}{({\bf m}^T {\bf e}_1)}\right)\,\left(\frac{1}{4}\,F^I\,\tilde{F}^J-2\,\Phi_2^I\,m^J\right)
-2\,\hat{\Phi}_1^I\,e_{2\,I}+2\,m^I\,e_{2\,I}\,,\label{newlagra}
\end{align}
where we used $C={\rm Tr}(\sqrt{\eta\,\tilde{\eta}_0})=2\lambda\,m^I\,e_{2\,I}$. By varying (\ref{newlagra}) with respect to $\hat{\Phi}_1$ and $\Phi_2$ we find  $
\hat{\Phi}_1^I=g^{-1\,IJ}\,e_{2\,J}$, $\Phi_2^I=g^{-1\,IJ}\,\theta_{JK}\,m^K$
which are just eq.s (\ref{solvesigma}) and (\ref{lambdasolved}) expressed in terms of the new auxiliary fields.
By  redefining $
\hat{\Phi}_1^I=-\Phi_1^I+m^I$
we may identify the F-terms of the ($\mathcal{N}=1$)-superfields as $Y^I\propto \Phi_1^I+i\,\Phi_2^I$ (the proportionality is intended through a real factor). We find
\begin{equation}
\frac{1}{\lambda}\left(C-f(\Lambda(\Phi))\right)=2\,{ m}^I\,{ e}_{2 I}-2\,\hat{\Phi}_1^I\,e_{2\,I}=2\, {\Phi}^I\,e_{2\,I}\,.
\end{equation}
This term combines with the following term in  (\ref{newlagra}):
\begin{equation}
2\,{\bf m}^T\,\left(\frac{{\bf e}_1 {\bf e}_1^T}{{\bf m}^T {\bf e}_1}\right)\,\Phi_2=2\,\Phi_2^I\,e_{1\,I}\,,\label{traceta}
\end{equation}
to form
$2\,\Phi_2^I\,e_{1\,I}+2\, {\Phi}_1^I\,e_{2\,I}\,\,\propto \,\,{\rm Im}\int d^2\theta\,e_I\,Y^I$,
where $e_I\equiv e_{1\,I}+i\,e_{2\,I}$. This is the chiral FI term of \cite{Ferrara:2014oka}.\par
If we vary (\ref{newlagra}) with respect to $z^i=X^I$ we find eq.s (\ref{eqsf}) written in terms of ${\Phi}_1$ and $\Phi_2$:
\begin{align}
C_{IJK}\left[-{i}\,\left({\Phi}_1^I{\Phi}_1^J+\Phi_2^I\,\Phi_2^J-2\,m^I \Phi_1^J-\frac{1}{4}\,F^IF^J\right)+
\left(\frac{F^I\tilde{F}^J}{4}-2\Phi_2^I\,m^J\right)\right]=0\,,\label{eqsf2}
\end{align}
which coincide with those found in \cite{Ferrara:2014oka}.\par

 A detailed analysis of the  $\mathcal{N}=2$ and of  the maximally extended $\mathcal{N}=4$ supersymmetric cases,
 is postponed to a forthcoming publication.

\section*{Acknowledgements}
We are particularly grateful to S. Ferrara for stimulating our interest in the subject and for enlightening discussions. We also thank A. Ceresole  and I. Pesando for useful discussions.

\end{document}